\documentclass[aps,prb,%
twocolumn,groupedaddress,superscriptaddress,amsfonts,amssymb,amsmath,citeautoscript,longbibliography,
10pt]{revtex4-2}

\usepackage[utf8]{inputenc}
\usepackage[english]{babel}

\usepackage{microtype} 
\usepackage{xspace} 

\usepackage{hyphenat}

\usepackage{txfonts}  
\usepackage{txfontsb} 

\usepackage{bm} 

\usepackage{xcolor}
\usepackage[]{graphicx} 
\graphicspath{{figs/}}

\usepackage[]{booktabs}
\usepackage{array}
\usepackage{multirow}

\usepackage{enumerate}
\usepackage[inline]{enumitem}

\usepackage{xr}
\makeatletter
\newcommand*{\addFileDependency}[1]{
  \typeout{(#1)}
  \@addtofilelist{#1}
  \IfFileExists{#1}{}{\typeout{No file #1.}}
}
\makeatother

\usepackage{hyperref}
\hypersetup{colorlinks,
	linkcolor={blue!75!black!80!yellow},
	citecolor={blue!75!black!80!yellow},
	urlcolor={blue!75!black!80!yellow}
}

\usepackage[capitalize,nameinlink]{cleveref}

\crefname{subequations}{Eqs.}{Eqs.} 
\Crefname{subequations}{Eqs.}{Eqs.}
\crefformat{subequations}{#2Eqs.~(#1)#3}
\Crefformat{subequations}{#2Eqs.~(#1)#3}
\crefname{page}{p.}{p.} 
\crefname{table}{Table}{Tables}
\crefname{figure}{Figure}{Figures}
\crefname{section}{Section}{Sections}

\usepackage{placeins}

\usepackage{siunitx}
\DeclareSIUnit[number-unit-product = ]\percent{\char`\%} 

\usepackage[centering,hmargin=18mm,tmargin=29.4mm,bmargin=24mm]{geometry}

\usepackage{soul}

\usepackage{textcomp} 
\usepackage{xifthen}
\usepackage{etoolbox}
\newboolean{togglecomments}
\newboolean{toggletodos}
\newboolean{togglechanges}

\setboolean{togglecomments}{true}
\setboolean{toggletodos}{true}
\setboolean{togglechanges}{false} 

\newcommand{\textblacksquare}{$\blacksquare$}
\newcommand{\todo}[1]{\ifbool{toggletodos}%
	{\textcolor{green!60!black}{\small\textsf{{}\textsuperscript{\textsc{\textsf{todo}}}}[\ignorespaces#1]}} 
	{}}     
\newcommand{\comment}[2]{\ifbool{togglecomments}%
		{\textcolor{blue!70!black}{\small\sf\textsuperscript{\textsc{\textsf{\ignorespaces#1}}}[\ignorespaces#2]}} 
		{}}     
\newcommand{\swap}[2]{\ifbool{togglechanges}
	{\ignorespaces#2}  
	{\textcolor{red!70!black}{[\ignorespaces#1]}\textrightarrow{}\textcolor{green!50!black}{[\ignorespaces#2]}}}
\newcommand{\remove}[1]{\ifbool{togglechanges}
	{}    
	{\textcolor{blue}{\ignorespaces#1}}}
\newcommand{\inset}[1]{\ifbool{togglechanges}
	{\ignorespaces#1}  
	{\textcolor{green!50!black}{\ignorespaces#1}}}

\newcommand{\citeremind}[1]{%
	[\textcolor{blue!75!black!80!yellow}{\textblacksquare%
		\ifthenelse{\isempty{#1}}{}{\textsuperscript{\tiny\textsf{\ignorespaces#1}}}%
	}]\xspace}

\newcommand{\appropto}{\mathrel{\vcenter{
			\offinterlineskip\halign{\hfil$##$\cr
				\propto\cr\noalign{\kern.2pt}\sim\cr\noalign{\kern-2.5pt}}}}}

\let\Re\relax 
\DeclareMathOperator{\Re}{Re}
\let\Im\relax 
\DeclareMathOperator{\Im}{Im}

\DeclareFontFamily{U}{mathx}{\hyphenchar\font45}
\DeclareFontShape{U}{mathx}{m}{n}{<5> <6> <7> <8> <9> <10>
                                  <10.95> <12> <14.4> <17.28> <20.74> <24.88>
                                  mathx10}{}
\DeclareSymbolFont{mathx}{U}{mathx}{m}{n}
\DeclareFontSubstitution{U}{mathx}{m}{n}

\makeatletter
\newcommand{\raisemath}[1]{\mathpalette{\raisem@th{#1}}}
\newcommand{\raisem@th}[3]{\raisebox{#1}{$#2#3$}}
\makeatother

\renewcommand{\paragraph}[1]{\vskip 1ex\noindent\textbf{#1.}~}

\usepackage[eulergreek]{sansmath}
\makeatletter
\renewcommand\@make@capt@title[2]{%
    \@ifx@empty\float@link{\@firstofone}{\expandafter\href\expandafter{\float@link}}%
    \sisetup{math-sf=\textsf}%
    \sansmath\sffamily\textbf{#1\@caption@fignum@sep}#2 
}%

\makeatother

\usepackage{physics}
\usepackage{orcidlink}

\usepackage{amsmath, amssymb}
\usepackage{import}
\usepackage{cleveref}
\usepackage{xcolor}
\usepackage{tikz}
\usetikzlibrary{positioning,decorations.pathreplacing,decorations.markings,math,calc}

\graphicspath{{}}

\setboolean{togglecomments}{true}
\setboolean{toggletodos}{true}
\setboolean{togglechanges}{false} 

\newcommand{\RPtwo}{\ensuremath{\mathbf{RP}^2}}
\newcommand{\KB}{{\ensuremath{\mathbf{K}^2}}}
\newcommand{\TO}{{\ensuremath{\mathbf{T}^2}}}
\newcommand{\Z}{\ensuremath{\mathbb{Z}}}
\newcommand{\R}{\ensuremath{\mathbb{R}}}
\newcommand{\C}{\ensuremath{\mathbb{C}}}
\newcommand{\B}{\ensuremath{\mathbb{B}}}

\renewcommand{\cp}{\ensuremath{{}_p}}
\newcommand{\cq}{\ensuremath{{}_q}}

\definecolor{dblu}{HTML}{162479}
\definecolor{lblu}{HTML}{266bb0}
\definecolor{lred}{HTML}{a83951}
\definecolor{dred}{HTML}{710022}

\begin{document}
\title{Exceptional Topology on Nonorientable Manifolds}

\author{J. Lukas K. König \orcidlink{0000-0002-1784-4619}}
\thanks{These authors contributed equally to this work}
\affiliation{Department of Physics, Stockholm University, AlbaNova University Center, 106 91 Stockholm, Sweden}
\author{Kang Yang \orcidlink{0000-0001-8329-1867}}
\thanks{These authors contributed equally to this work}
\affiliation{Dahlem Center for Complex Quantum Systems and Fachbereich Physik, Freie Universität Berlin, 14195 Berlin, Germany}
\author{Andr\'e Grossi Fonseca \orcidlink{0000-0001-9789-8037}}
\affiliation{Department of Physics, Massachusetts Institute of Technology, Cambridge, Massachusetts 02139, USA}
\author{Sachin Vaidya \orcidlink{0000-0002-3569-6568}}
\affiliation{Department of Physics, Massachusetts Institute of Technology, Cambridge, Massachusetts 02139, USA}
\author{Marin Solja\v{c}i\'c \orcidlink{0000-0002-7184-5831}}
\affiliation{Department of Physics, Massachusetts Institute of Technology, Cambridge, Massachusetts 02139, USA}
\author{Emil J. Bergholtz \orcidlink{0000-0002-9739-2930}}
\email{emil.bergholtz@fysik.su.se}
\affiliation{Department of Physics, Stockholm University, AlbaNova University Center, 106 91 Stockholm, Sweden}
\date{March 1, 2025}

\begin{abstract}
We classify gapped phases and characteristic nodal points of non-Hermitian band structures on two-dimensional nonorientable parameter spaces.
Such spaces arise in a wide range of physical systems in the presence of nonsymmorphic parameter space symmetries.
For gapped phases, we find that nonorientable spaces provide a natural setting for exploring fundamental structural problems in braid group theory, such as torsion and conjugacy. 
Gapless systems, which host exceptional points (EPs), explicitly violate fermion doubling, even in two-band models. 
We demonstrate that EPs traversing the nonorientable parameter space exhibit non-Abelian charge inversion.
These braided phases and their transitions leave distinct signatures in the form of bulk Fermi arc degeneracies, offering a concrete route toward experimental realization and verification.
\end{abstract}
\maketitle

\section{Introduction} 
    
\begin{figure}
\fboxsep=1pt
    \newcommand{\btotKB}{\ensuremath{
        ~B_\text{tot}^\KB\!
        =
        \bbdyKa
        \bbdyKb
        \!\left(\bbdyKa\right)^{-1}\!\!
        \bbdyKb}}
    \newcommand{\btotRP}{
        ~~\ensuremath{B_\text{tot}^{\RPtwo}=
        \left(\bbdyRP\right)^2}}
    \newcommand{\bbdyRP}{\textcolor{dred}{\ensuremath{B_{pq}^{\RPtwo}}}}
    \newcommand{\bbdyKb}{\textcolor{lred}{\ensuremath{B_p^\KB}}}
    \newcommand{\bbdyKa}{\textcolor{lblu}{\ensuremath{B_q^\KB}}}
    \centering
    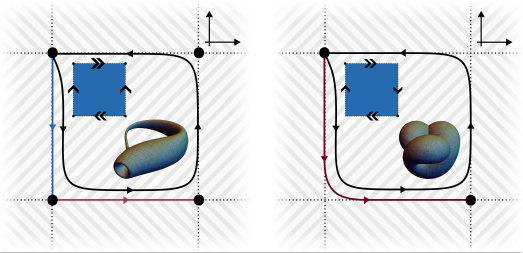
    \caption{
    Exceptional phases on nonorientable manifolds: Under the boundary identification shown in the top-left insets, a free parameter space becomes (a) the Klein bottle, (b) the real projective plane, illustrated as an immersion into \(\R^3\) in the bottom-right insets.
    The red and blue paths are different types of noncontractible loops in the space. They can carry nontrivial eigenvalue braids and play central roles in the classification.
    The black loop enclosing the fundamental domain is a combination of these fundamental loops. 
    It is contractible and imposes constraints on the braids along the noncontractible loops for gapped phases.
    This black loop also determines the fermion doubling question for gapless phases.
    Background shading denotes the orientation of the neighboring unit cells. 
    }
    \label{fig:brillouin-zones}
\end{figure}

Non-Hermitian phases have been broadly studied in systems where gain and loss are non-negligible, which range from condensed matter to optics, acoustics, and electrical circuits \cite{berryPhysicsNonhermitianDegeneracies2004, moiseyevNonHermitianQuantumMechanics2011, zhuPTSymmetricAcoustics2014, zhouObservationBulkFermi2018, cerjanExperimentalRealizationWeyl2019,ashidaNonHermitianPhysics2020,helbigGeneralizedBulkBoundary2020,pangSyntheticNonAbelianGauge2024,qiuMinimalNonabelianNodal2023,qiuMinimalNonabelianNodal2023,wangExceptionalNexusBoseEinstein2024, zhangExperimentalCharacterizationThreeband2023}. 
Non-Hermitian bands and their degeneracies, namely exceptional points (EPs) \cite{katoPerturbationTheoryLinear1995, dingNonHermitianTopologyExceptionalpoint2022,miriExceptionalPointsOptics2019, liExceptionalPointsNonHermitian2023, zhenSpawningRingsExceptional2015, hodaeiEnhancedSensitivityHigherorder2017,  yoshidaWindingTopologyMultifold2025, dingEmergenceCoalescenceTopological2016}, often challenge cornerstones of topological physics, such as the fermion doubling theorem \cite{nielsenAbsenceNeutrinosLattice1981, nielsenAbsenceNeutrinosLattice1981a, sunAliceStringsNonHermitian2020,
yangFermionDoublingTheorems2021,wojcikEigenvalueTopologyNonHermitian2022,huKnotTopologyExceptional2022,konigBraidprotectedTopologicalBand2023} and the bulk-boundary correspondence \cite{leeAnomalousEdgeState2016,yaoEdgeStatesTopological2018,kunstBiorthogonalBulkBoundaryCorrespondence2018,bergholtzExceptionalTopologyNonHermitian2021,okumaNonHermitianTopologicalPhenomena2023}.
This exceptional topology is closely tied to the study of braids and knots \cite{renBraidGroupTopological2013,shenTopologicalBandTheory2018,wojcikHomotopyCharacterizationNonHermitian2020,liHomotopicalCharacterizationNonHermitian2021,
huKnotsNonHermitianBloch2021,wangTopologicalComplexenergyBraiding2021,zhangObservationAcousticNonHermitian2023,yangHomotopySymmetryNonHermitian2024}, formed by complex spectra in non-Hermitian phases winding around each other. On orientable parameter spaces, the non-Abelian character of the braid group has revealed many interesting physically observable consequences, including path-dependent fusions of EPs and spectral flows \cite{papNonabelianNatureSystems2018,patilMeasuringKnotNonHermitian2022,guoExceptionalNonAbelianTopology2023,zhongNumericalTheoreticalStudy2023,fuBraidingTopologyNonHermitian2024,guriaResolvingTopologyEncircling2024,longNonAbelianBraidingTopological2024,wangObservationBraidProtectedUnpaired2026}. These observations result from the interplay between the braid group and the underlying topology of the orientable parameter space, typically a torus or a disk.

Recent work on Hermitian systems has shown that more exotic parameter spaces are physically accessible, and that they fundamentally change the topological classification \cite{chenBrillouinKleinBottle2022,zhangGeneralTheoryMomentumSpace2023,grossiefonsecaWeylPointsNonorientable2024, vaidya2025response, shangObservationHigherOrderEnd2024, morimotoFloquetTopologicalPhases2017, huHigherOrderTopologicalInsulators2024, wangChessboardAcousticCrystals2023}. 
For example, certain configurations of gauge fluxes \cite{chenBrillouinKleinBottle2022, liKleinbottleQuadrupoleInsulators2023,puAcousticKleinBottle2023,wangChessboardAcousticCrystals2023,taoHigherorderKleinBottle2024, zhuBrillouinKleinSpace2024,huHigherOrderTopologicalInsulators2024,huHigherOrderTopologicalInsulators2024,huAcousticHigherorderTopological2025,nguyenFermiArcReconstruction2023,laiRealprojectiveplaneHybridorderTopological2024,shenObservationKleinBottle2025,wangOpticalInterfaceStates2017,qiuOctupoleTopologicalInsulating2025}, spin-space groups \cite{xiaoSpinSpaceGroups2024}, or moiré structures \cite{calugaruMoireMaterialsBased2025} in real-space translate into nonsymmorphic symmetries in momentum space and render the Brillouin zone a nonorientable manifold. 
This enriches topological physics, reducing Chern numbers to $\mathbb{Z}_2$ symmetry-protected invariants~\cite{chenBrillouinKleinBottle2022}, and modifies fermion doubling, requiring even, but possibly nonvanishing total chirality~\cite{grossiefonsecaWeylPointsNonorientable2024, zhangBrillouinPlatycosmsTopological2025}.

Such parity requirements are characteristic of Abelian topology on nonorientable spaces. 
As braids and knots on nonorientable surfaces differ decidedly from those on tori or disks, one may naturally expect richer symmetry-protected non-Hermitian phases and unique, distinctly non-Abelian features to arise in the presence of non-Hermiticity and nonorientability.
In this work, we explore this intersection by investigating the exceptional topology in nonorientable spaces, and its implications for gap closing.
We first apply the theory of braided band structures to nonorientable parameter spaces, using as examples the Klein bottle \KB{} and the real projective plane \RPtwo. 
The classification of gapped bands is given by eigenvalue braids satisfying certain conditions along noncontractible loops in these nonorientable parameter spaces, while for gapless systems, the summation of topological charges of all EPs must respect specific forms.
Then, we discuss phase transitions, which occur when EPs encircle the parameter space.
We provide concrete rules how the encircling EP and the topological phase affect each other.
Finally, we show gapless systems can exhibit a kind of unpaired monopole that is forbidden both in Hermitian systems and in non-Hermitian systems on orientable spaces.

\section{Review of Exceptional Braid Topology}\label{sc_EBT}
We start by describing how mutual windings, or braids, of complex eigenvalues characterize non-Hermitian systems \cite{shenTopologicalBandTheory2018,liHomotopicalCharacterizationNonHermitian2021,wojcikHomotopyCharacterizationNonHermitian2020,huKnotsNonHermitianBloch2021,yangHomotopySymmetryNonHermitian2024}.
Such systems are described by a $m \times m$ non-Hermitian matrix $H(\lambda)$, with $\lambda$ a set of parameters. The eigenvalues of $H(\lambda)$ are complex numbers $E_1(\lambda),E_2(\lambda)\dots ,E_m(\lambda)$.
If there is no EP at $\lambda$, these complex eigenvalues are distinct, and the complex level spacings $\Delta_{ij}(\lambda)=E_i(\lambda)-E_j(\lambda)$ are nonzero. For an $m$-band system, we can follow a parametric loop in $\lambda$: $\lambda(t+2\pi)=\lambda(t)$. If no EP is encountered in this $\lambda$-loop, the $m$ distinct complex eigenvalues \(E_i\) rotate around each other in the complex plane and then return to their initial configuration \cite{zhongNumericalTheoreticalStudy2023}. 
Their trajectories, plotted as a function of $t$, trace out $m$ strands that wind around each other. Such mutually winding $m$ strands determine a braid pattern $B$, which is an element of the Artin braid group \(\B_m\). 

The braid pattern of $m$ noncolliding strands is a topological invariant: it cannot change under continuous deformation of the $\lambda$-loop until a level crossing is met where some \(E_j(\lambda)=E_k(\lambda)\) \cite{kasselBraidGroups2008}. Therefore, by taking a $\lambda$-loop to encircle EPs in parameter space, the braid of complex eigenvalues gives the topological charge of these EPs \cite{shenTopologicalBandTheory2018,liHomotopicalCharacterizationNonHermitian2021,wojcikHomotopyCharacterizationNonHermitian2020,huKnotTopologyExceptional2022}. The result of merging two EPs with braid charges $B,B'$ can be read out by looking at their combined charge $BB'$, where the product of two braids is obtained by concatenating their strand's windings \cite{wojcikEigenvalueTopologyNonHermitian2022,konigBraidprotectedTopologicalBand2023}.

We can also use the above eigenvalue braid characterization to describe gapped bands \cite{shenTopologicalBandTheory2018,wojcikHomotopyCharacterizationNonHermitian2020,liHomotopicalCharacterizationNonHermitian2021,yangHomotopySymmetryNonHermitian2024}. 
By considering the complex level spacings as a generalization of real-valued band gap, a set of bands is \emph{gapped} if their level spacings are nonzero everywhere \cite{shenTopologicalBandTheory2018,wojcikHomotopyCharacterizationNonHermitian2020,liHomotopicalCharacterizationNonHermitian2021,yangHomotopySymmetryNonHermitian2024}:
\begin{equation}
    \abs{\Delta_{ij}(\lambda)}=\abs{ E_i(\lambda)-E_j(\lambda)} >0, \textrm{ for all }i,j,\lambda.
\end{equation}
For gapped bands, all parametric loops in $\lambda$ give braid patterns of the complex eigenvalues. 
If we can gradually contract the loop to a single point in the $\lambda$-space, the $m$ strands traced out during the $\lambda$-loop become straight lines (constant values) after the contraction, leading to a trivial braid \(1\). Thus, contractible loops in the $\lambda$-space always correspond to trivial braid elements for gapped bands.
In gapped bands, nontrivial braids appear when we follow the evolution of the complex eigenvalues along noncontractible loops in the $\lambda$-space. For example, if the parameter space of $\lambda$ is a torus, such as the two-dimensional Brillouin zone $\lambda=(k_x,k_y)\in[-\pi,\pi]\times[-\pi,\pi]$ with identified opposite boundaries, the two loops $(k_x,k_y)=(t,\textrm{constant})$ and $(k_x,k_y)=(\textrm{constant},t)$ parametrized for \(t\in[-\pi,\pi]\) are noncontractible. 
For each of them we can identify a braid element, denoted as $ B_x, B_y$ respectively. 
Even though each large loop on the torus is noncontractible, their combination, $(k_x,k_y)\to (k_x,k_y+2\pi)\to (k_x+2\pi,k_y+2\pi)\to(k_x+2\pi,k_y)\to(k_x,k_y)$, is contractible because this loop encloses a first Brillouin zone and thus can be continuously deformed into a point (constant loop). This tells us that on a torus, we have the constraint $B_xB_yB^{-1}_xB^{-1}_x=1$ for gapped bands.

From the above torus example, we see how to classify the exceptional topology of gapped bands: 
\textbf{(i)} We identify all noncontractible loops in the parameter $\lambda$-space. Each type of noncontractible loop can give rise to a nontrivial braid of the complex eigenvalues of $H(\lambda)$. \textbf{(ii)} We check whether some combination of noncontractible loops becomes contractible. For each combination, we put a constraint on the braids obtained in the first step. The exceptional topology is characterized by the nontrivial eigenvalue braids and the constraints that the braids must satisfy. \textbf{(iii)} Since the braid group is non-Abelian (for $m>2$), there is another redundancy in applying the braid group to classify the topology \cite{merminTopologicalTheoryDefects1979}. The braid pattern explicitly depends on the position of the starting (or base) point $\lambda(0)$, while the topological classes of gapped bands are global properties of the mutually winding patterns and therefore independent of $\lambda(0)$.
This redundant base-point dependence requires us to identify all braid patterns $B$ in its conjugacy class \cite{merminTopologicalTheoryDefects1979}, given by $bBb^{-1}$  where $b$ ranges over all elements in the braid group. 
For the previous example of the torus $\lambda$-space, if we label the braids along the $k_x$- and the $k_y$-direction as $(B_x,B_y)$, we need to place the gapped bands characterized by $(B_x,B_y)$ and the gapped band characterized by $ (bB_xb^{-1},bB_yb^{-1})$ \cite{wojcikHomotopyCharacterizationNonHermitian2020,liHomotopicalCharacterizationNonHermitian2021} into the same topological class. 

In sum, on a torus, the exceptional topology is classified by  the braid pair $(B_x,B_y)$ [step (i)] satisfying $B_xB_yB^{-1}_xB^{-1}_x=1$ [step (ii)], and all pairs related by the conjugation relation $(B'_x,B'_y)=(bB_xb^{-1},bB_yb^{-1})$ are identified as the same class [step (iii)]. We will use these three steps to classify the exceptional topology of gapped bands on $\lambda$-spaces consisting of Klein bottles and real projective planes in Sec.~\ref{sc_not}. Explicit examples will be given in Sec.~\ref{sc_eggbands}. In the supplemental material we provide a more mathematical derivation of our results encompassing all two-dimensional closed surfaces, using established results from topology \cite{footnoteSupplemental}. 
Since step (iii) does not rely on the properties of the $\lambda$-space, we will only explicitly show the results of steps (i) and (ii).

On the other hand, when there are EPs in the system, we may wonder what happens when all EPs are brought together. This is closely related to the fermion doubling question \cite{nielsenAbsenceNeutrinosLattice1981}. The ways of bringing EPs together on a closed surface is not unique. The results may depend on the topology of the trajectories of the EPs~\cite{konigBraidprotectedTopologicalBand2023}, which is a typical feature of point defects carrying non-Abelian charges \cite{merminTopologicalTheoryDefects1979}. A convenient way to consider merging of EPs is to restrict all EPs to move only within a disklike region. In this situation, by looking at the topological charge measured from the boundary loop of the disk, we know the results of bringing all EPs together within the disk. Since we often have a finite number of EPs, the disklike region can be chosen to bound the whole parameter space. In this situation, the topology of the parameter space will restrict the possible total charges that EPs can have. We will discuss these questions in Sec.~\ref{sc_not} and Sec.~\ref{sc_epmp}.

To express the braid group more conveniently, we will write the braid elements in terms of the products of simpler generators \(\sigma_j\in \B_m\) and their inverse in the braid group. These are braids that  intertwine two adjacent strands counterclockwise in the complex plane \cite{artinTheorieZoepfe1925}.  
For example, $\sigma_j$ exchanges neighboring eigenvalues \(E_j,E_{j+1}\) counterclockwise in the complex plane while keeping all other eigenvalues invariant.

\section{Braids in nonorientable spaces}\label{sc_not}

\subsection{Nonorientable parameter spaces}
We now consider a parameter space whose geometry is equivalent to a two-dimensional nonorientable manifold, parametrized by two coordinates $\lambda=(p,q)$. These coordinates can represent (quasi-)momenta \cite{wangTopologicalComplexenergyBraiding2021,liExceptionalPointsNonHermitian2023}, or more generic parameters \cite{ozturkObservationNonHermitianPhase2021,bergmanObservationAntiparitytimesymmetryPhase2021,liExceptionalPointsNonHermitian2023,miriExceptionalPointsOptics2019} depending on the physical platform. 
In this work we use \(p, q\) to denote general parameters, regardless of physical origin. As we will see, such nonorientable manifolds can appear as momentum-space fundamental domains, i.e., parts of a two-dimensional Brillouin zone encoding all information of the system. For this reason, we also call these nonorientable manifolds fundamental domains. 
While our subsequent results hold in general for nonorientable spaces, we now elaborate on this specific mechanism.

Starting from an originally \(2\pi\)-periodic matrix \(H(p,q) = H(p+2\pi,q) = H(p,q+2\pi)\) we impose a nonsymmorphic parameter space symmetry of the form \cite{chenBrillouinKleinBottle2022} 
\begin{equation}
\label{eq:nonsymmorphic}
    H(p,q) = H(-p,q+\pi).
\end{equation}
As this glide symmetry identifies half of the $(p,q)$-space to the other half (shifted by $q+\pi$), we can restrict our study to the fundamental domain \([-\pi,\pi]\times[0,\pi]\), where the symmetry acting on the boundaries gives rise to the \KB{} topology.

In a similar manner, we can obtain the real projective plane \RPtwo{} as a fundamental domain, by imposing a second glide symmetry
\begin{equation}
    H(p,q) = H(p+\pi,-q)
\end{equation}
in addition to \cref{eq:nonsymmorphic}.
Under the combined action of these two symmetries, we can further reduce the fundamental domain by half to \([-\pi/2,\pi/2]\times[-\pi/2,\pi/2]\). All boundaries of this square are then identified with the opposing side with reversed orientation, leading to \RPtwo{} 
topology for the fundamental domain \cite{footnoteSpacegroup,grossiefonsecaWeylPointsNonorientable2024}.
Some care must be taken regarding the corner points of the fundamental domain.
They are fixed points, i.e., invariant under the combined action of both symmetries, which transforms \((p,q) \to (\pi-p,\pi-q)\).
The local neighborhood of these points is rotated by a half-turn. 
Invariance of the Hamiltonian under both symmetries, therefore, means that these points are pairwise identified with the opposite corner, and that the Hamiltonian locally must be invariant under a half-turn rotation group \(C_2\).
In Hermitian systems such a constraint can make degeneracies more abundant: the Hamiltonian for any two-bands which transform differently under the rotation must be of the form \(H= c \sigma_z\) (ignoring constant eigenvalue offset), for \(\sigma_z\) a Pauli matrix and \(c\) a real constant. 
The two bands become degenerate when \(c=0\), i.e., with only one tuning parameter. 
Nevertheless, rotations do not protect degenerate points in the same way in non-Hermitian systems \cite{konigNodalPhasesNonHermitian2024}, so these two points are not special compared to other $(p,q)$-points and do not need to be treated independently for the purposes of eigenvalue topology.

\subsection{Braid relations and Abelianization}

First, we classify the exceptional topology for gapped bands. The key step is to find the noncontractible loops of the fundamental domain. To visualize different types of loops on Klein bottles and real projective planes, it is convenient to draw these spaces as rectangles and glue the edges of the rectangle (see \cref{fig:brillouin-zones}). Different types of noncontractible loops can be readily identified by looking at the rectangle's boundaries, starting from a corner point. According to step (iii), after replacing the braids by their conjugation classes, the classification is independent of the choice of base point.

From \cref{fig:brillouin-zones}, for the Klein bottle, all four corner points of the rectangle are identified as the same point after the edges are glued to form the Klein bottle. Therefore, similar to the torus, the vertical and horizontal edges (marked in blue and red respectively) are noncontractible loops, carrying nontrivial complex eigenvalue braids $(B_q^\KB,B_p^\KB)$. For the projective plane, only the pairs of corner points across the diagonals of the rectangle are identified. The edges of the rectangle are arcs, with a noncontractible loop found by connecting the horizontal and vertical edges together, with braid \(B_{pq}^{\RPtwo}\). Unlike the torus and the Klein bottle, there is only one type of noncontractible loop on the real projective plane.

So far we have carried out step (i) of the procedure listed in Sec.~\ref{sc_EBT}. We now need to find out the constraints on the braid invariants on these nonorientable spaces. 
To carry out step (ii), we note that there is a contractible loop enclosing the fundamental domain, shown in black in \cref{fig:brillouin-zones}.
We identify the sequence of noncontractible loops (red/blue in \cref{fig:brillouin-zones}) that combines to this black loop. 
To do so, we must take into account that opposing boundaries of the fundamental domain are identified (potentially with reversed orientation), and therefore correspond to the same (reversed) loops.
These combined loops must then carry a trivial braid, since the combination is contractible. 
On the spaces we discuss explicitly here, we find that the combinations 
\begin{align}
    B_q^\KB 
    B_p^\KB
    \left(B_q^\KB\right)^{-1} 
    B_p^\KB=1, \quad \quad 
    \left(B_{pq}^{\RPtwo}\right)^2=1.\label{eq_tpconstr}
\end{align}
have to vanish.

The above relations are very different from those on the torus. There, the exceptional topology is characterized by two mutually commuting elements from the braid group $\{(B_x,B_y)|B_xB_yB_x^{-1}B_y^{-1}=1\}$. This commutation constraint is trivial for Abelian groups, while nontrivial for non-Abelian groups. The constraint on the Klein bottle on the other hand, is not a commutator, and therefore also nontrivial for topology characterized by Abelian groups. 
The constraint on the real projective plane is yet of another type; since there is only one type of noncontractible loop, the constraint is on a power of $B_{pq}^{\RPtwo}$, the braid element corresponding to this loop.

In the next section, we will look for explicit examples of gapped topological bands. We need to solve the constraints \cref{eq_tpconstr}. For a non-Abelian group, which is the case of  braid groups $\mathbb B_m$ with $m > 2$, such constraints can be very complicated to solve. In order to have a simpler criterion before solving the full non-Abelian group relations, we introduce the Abelianization $\mathbb B_m/[\mathbb B_m,\mathbb B_m]$ of the braid group \cite{kasselBraidGroups2008,konigBraidprotectedTopologicalBand2023,yangHomotopySymmetryNonHermitian2024,shenTopologicalBandTheory2018,yangFermionDoublingTheorems2021}.
These are weaker versions of the constraints in \cref{eq_tpconstr}, except for the case of two-band models, where the braid group \(\B_2\) on two strands itself is Abelian and equivalent to its Abelianization.
Given some braid \(B\), its Abelianization \(A\in\Z\), also known as the degree or vorticity of \(B\), is the difference between its number of overcrossings \(\sigma_j\) and undercrossings \(\sigma_j^{-1}\). It can be explicitly computed by $A=1/\pi \sum_{i<j}\oint d\lambda\,\partial_\lambda\textrm{arg}(E_i-E_j)$ \cite{yangHomotopySymmetryNonHermitian2024,yangFermionDoublingTheorems2021} (not to be confused with the reference-point-gap winding, which usually gives a different value).
As mentioned before, Abelianizing the torus constraint $B_xB_yB_y^{-1}B_x^{-1}$ gives a trivial identity. 
In comparison,
the Abelianized constraints on nonorientable spaces are not trivial, giving
\begin{equation}
     A_q^\KB + A_p^\KB - A_q^\KB + A _p^\KB = 2A _p^\KB=0,
    \quad  
    2 A_{pq}^{\RPtwo}=0.\label{eq_srab}
\end{equation}
Here, as is common convention, we use $+$ for products of elements in an Abelian group. 

Now we consider gapless systems and study what happens when all EPs in a gapless phase are brought together. According to the results in Sec.~\ref{sc_EBT}, the total topological charge of the EPs within a disk is given by the complex eigenvalue braid when we go around the disk boundary. This total charge determines what happens when all EPs are moved to the same position inside the disk. As the eigenvalue braid does not change if we expand the disk without encountering an EP, we can deform the disk to bound the whole parameter space, such as the rectangles in \cref{fig:brillouin-zones}. As the boundary of the rectangle decomposes into the noncontractible loops as discussed above, this indicates that the total topological charge of all EPs inside the rectangle should have a decomposition:
\begin{equation}
    \prod_{\textrm{all EPs}} B^\KB_{\textrm{EP}}=B_q^\KB 
    B_p^\KB
    \left(B_q^\KB\right)^{-1} 
    B_p^\KB,\quad \prod_{\textrm{all EPs}} B^{\RPtwo}_{\textrm{EP}}=\left(B_{pq}^{\RPtwo}\right)^2.\label{eq_sumchg}
\end{equation}
Notice that here we use product symbols instead of summation symbols because the braids are in general non-Abelian. The existence of these decompositions implies that the total charges of all EPs brought together must take some specific form, which will decide whether the EPs annihilate, leading to a gap opening, or fuse into a single degeneracy. Similarly to the constraints for gapped bands in \cref{eq_srab}, we can also Abelianize the relation \cref{eq_sumchg}. This gives
\begin{equation}
    \sum_{\textrm{all EPs}} A^\KB_{\textrm{EP}}=2 
    A_p^\KB,\quad \sum_{\textrm{all EPs}} A^{\RPtwo}_{\textrm{EP}}=2A_{pq}^{\RPtwo}.\label{eq_absumchg}
\end{equation}
We can see for these nonorientable surfaces, the total number of mutual windings between the complex eigenvalues is an even integer, and in general need not vanish. This is reminiscent of the Hermitian case, where only the parity of the Chern number is physically meaningful on nonorientable spaces \cite{chenBrillouinKleinBottle2022}, which leads to a $\mathbb{Z}_2$ fermion doubling theorem \cite{grossiefonsecaWeylPointsNonorientable2024}.
Since for the two-band case the Abelianization is equivalent to the original braid, this constraint is the full picture for two-band models.
For three or more bands the full non-Abelian description in terms of braids is necessary, since there exist nontrivial braids with Abelianization zero, such as \(\sigma_2 \sigma_1^{-1}\).

The similarity between equations \cref{eq_tpconstr} of the braid constraints for gapped bands and equations \cref{eq_sumchg} of the total EP charge for gapless bands is not accidental. In fact, it is related to a systematic way of constructing compact surfaces from disks of different dimensions, known as the cell complex construction \cite{munkres2018elements}.

\section{Gapped Phases and Fermi Arcs}\label{sc_eggbands}

\begin{figure}
    \centering
    \newcommand{\bx}{\textcolor{lred}{\ensuremath{B_p}}}
    \newcommand{\tbx}{\textcolor{dred}{\ensuremath{\tilde{B}_p}}}
    \newcommand{\bep}{\textcolor{lblu}{\ensuremath{B_\text{EP}}}}
    \newcommand{\tbep}{\textcolor{dblu}{\ensuremath{\tilde{B}_\text{EP}}}}
    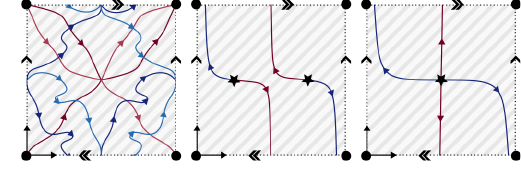
    \caption{
    Fermi arcs as experimentally accessible signatures.
    Real (imaginary) Fermi arcs as defined in \cref{eq:fermi-arc-condition} are drawn in red (blue), with the orientation defined in the main text.
    (a) The gapped three-band model on \KB{} given in \cref{eq:klein-model-gapped}, with lighter and darker tones denoting Fermi arcs of different band pairs.
    The exceptional phase, identified by the braids \((\sigma_2\sigma_1^{-1},\Delta^{-1})\) along the $p-$ and $q-$directions, has three Fermi arcs encircling the fundamental domain in the positive \(p\)-direction, originating from the Abelianization of \(\Delta\). 
    This clearly distinguishes this phase from the trivial flat-band phase \((1,1)\).
    (b) The two-band model in \cref{eq:model-two-band} at \(\ell=0\), perturbed to \(e^{0.02i} H^\KB_\text{EP}\) to better showcase the Fermi arcs.
    A real and an imaginary Fermi arc emerge from each EP (star marker), with a direction given by the eigenvalue crossing behavior. This direction is inverted on the nonorientable boundary, allowing for a total flux out of the fundamental domain that matches the total braid degree, \(2\).
    (c) The two-band model in \cref{eq:model-two-band}, perturbed as in (b) and tuned to \(\ell=1\) exhibits a characteristic monopole.
    Two real and two imaginary Fermi arcs emerge from the monopole, accounting for its total braid degree \(\sum A_\text{EP}^\KB=2\). Such an EP cannot exist alone on an orientable space, where outgoing Fermi arcs cannot be compensated.
    }
    \label{fig:branch-cuts}
\end{figure}

\subsection{General statements}
We now leverage the braid relations derived in the previous section to fully characterize gapped phases, i.e., phases without EPs, where the gaps \(\Delta_{jk}=E_j-E_k\) are nonzero throughout parameter space. 
As we showed, these phases are characterized by the respective fundamental braids for the given parameter space, e.g., the pair \(\left(B_p^{\KB}, B_q^{\KB}\right)\) on the two fundamental loops of the Klein bottle, and \(B_{pq}^{\RPtwo}\) on the single fundamental loop of the real projective plane.
These braid elements have to satisfy the constraints \cref{eq_tpconstr}.

We begin with the simpler relation, that on the real projective plane. A gapped phase on \RPtwo{} must satisfy \(1=\left(B_{pq}^{\RPtwo}\right)^2\) .
This condition amounts to finding braid elements that square to unity (i.e., have $\mathbb Z_2$ torsion in group theory language). Since the braid group is torsion-free \cite{kasselBraidGroups2008}, the choice
\begin{equation}
	B_{pq}^{\RPtwo} = 1
\end{equation}
is the only solution.
Hence, there is only one trivial exceptional phase for gapped bands on \RPtwo{}.

Gapped phases on the Klein bottle need to satisfy the first equation in \cref{eq_tpconstr}. It is more convenient to  express the constraint in the form
\begin{equation}
\label{eq:klein-gapped}
    B_p^\KB = B_q^\KB \left(B_p^\KB\right)^{-1} \left(B_q^\KB\right)^{-1},
\end{equation}
i.e., one braid \(B_p^\KB\) must be conjugate to its own inverse, with another braid \(B_q^\KB\) as the conjugating element. If we Abelianize this equation,
the requirement reads \(A_p^\KB = 0\), i.e., zero braid degree along the direction parallel to \(p\). Since in two-band models the braid group $\mathbb B_2=\mathbb Z$ itself is Abelian, two-band gapped phases must thus have trivial \(B_p^\KB = 1\), and are solely classified by a free choice of \(B_q^\KB\), which is indexed by how many times the two eigenvalues wind around each other.

For systems of three or more bands, we need to solve the non-Abelian conjugacy question, \cref{eq:klein-gapped}, in full, which is a key braid group structure and usually very complicated. 
An example of a nontrivial solution is \(B_p^\KB = \sigma_2 \sigma_1^{-1}\).
This braid is indeed conjugate to its own inverse, i.e., it satisfies \cref{eq:klein-gapped} for the choice  \(B_q^\KB=\Delta=\sigma_1\sigma_2\sigma_1\). 
Here, \(\Delta\) is the fundamental braid on three strands, which plays an important role in the study of braid groups \cite{kasselBraidGroups2008}. 
Conjugation with \(\Delta\) is equivalent to flipping a braid upside down, i.e., replacing \(\sigma_1\leftrightarrow\sigma_2\), \(\sigma_1^{-1}\leftrightarrow\sigma_2^{-1}\), which transforms the choice of \(B_p^\KB\) given here into its inverse.
In the supplemental material, we discuss \cref{eq:klein-gapped} from a group theory perspective using known conjugacy results for braid groups, and enumerate the list of gapped phases in three bands explicitly \cite{footnoteSupplemental}.

In stark contrast to classifying gapped phases on the torus, which is solved by mutually commuting braids, gapped phases on nonorientable surfaces bring us to two fundamental structural questions in braid group study, torsion and conjugacy class.

A different way to compare the two spaces is to construct a torus out of two copies of \KB, by gluing them along the nonorientable boundary. 
In the setting of \cref{eq:nonsymmorphic}, we can equivalently understand this as breaking the nonsymmorphic symmetry, and thus needing to describe the given system on the full torus rather than on two mirrored copies of \KB.
The fundamental braids on the torus can then be obtained from the \KB{} braids on the half-torus as 
\(B^{\mathbf{T}^2}_p = B^{\KB}_p\) and \(B^{\mathbf{T}^2}_q = (B^{\KB}_q)^2\).
In that scenario, all gapped phases on \KB{} are also gapped on the torus.
Conversely, there exist gapped phases on the torus that cannot be brought into gapped phases on \KB{} by imposing a symmetry, such as all phases with nonzero degree in \(B_p\) and trivial or commuting \(B_q\).

\subsection{Braided phase: three-band example}

A model that realises the Klein bottle gapped phase  \((\sigma_2\sigma_1^{-1},\Delta^{-1})\) is given by
\begin{equation}
  \label{eq:klein-model-gapped}
  H^\KB_\text{gap}=\mqty(
        -a &  1 & 0 \\
        \hphantom{-}1 & b & 1 \\
        \hphantom{-}0 & 1 & a
    ),\qq{where} \begin{cases}
      ~a&=i - i e^{2 i q} \cos{p},
      \\
      ~b&=(1+i) ~ e^{3 i q}\sin{p},
    \end{cases}
\end{equation}
in a parameter space that has been experimentally accessed in an acoustic setup \cite{tangExceptionalNexusHybrid2020}.

Such braided phases can be detected experimentally by their Fermi arcs. 
In gapped phases, albeit the complex band gaps \(\Delta_{jk}=E_j-E_k\ne 0\) are nonzero, their real parts and imaginary parts can vanish individually. Fermi arcs are the lines in parameter space where 
\begin{equation}
\label{eq:fermi-arc-condition}
    \Re\Delta_{ij}=\Re(E_i-E_j) = 0, ~~\text{or}~~ \Im\Delta_{ij}=\Im(E_i-E_j) = 0,
\end{equation}
for some pair of eigenvalues \(E_i, E_j\) (see \cref{fig:branch-cuts}). 
These Fermi arcs can be measured directly, for example in photonic crystals \cite{zhouObservationBulkFermi2018}, and thus form an accessible signature of gapped phases.
Fermi arcs in phases with nontrivial braids are irremovable: As a braid \(\sigma_i\) swaps two eigenvalues along the direction $p$ or $q$, both a real and imaginary Fermi arc must be present. 
Without fine-tuning, and assuming the system depends differentiably on its parameters, Fermi arcs form continuous lines that can only terminate at EPs.
In gapped phases (without EPs), Fermi arcs therefore generically form closed loops in parameter space.

Fermi arcs can be oriented consistently, such that eigenvalues swap places clockwise when crossing the Fermi arc left-to-right.
This corresponds to distinguishing crossing strands \(\sigma_i\) from the inverse operation. 
Signed counting of oriented Fermi arcs along a loop thus allows access to a braid's degree.
The orientation reversal across a nonorientable boundary implies that a Fermi arc's orientation is reversed on crossing.
In order to form closed loops with a consistent orientation, Fermi arcs in gapped phases must therefore cross nonorientable boundaries an even number of times, in similarity to 3D gapless Hermitian systems \cite{grossiefonsecaWeylPointsNonorientable2024}.
This is also reflected in the fact that the corresponding braid degrees in \cref{eq_srab} must be trivial. 

We note that experimentally determining which eigenvalues cross at a Fermi arc, together with its orientation, allows one to single out a braid group element \(\sigma_i^{\pm1}\), and thus completely reconstruct the full braid group description. Therefore, enforcing the stricter gap condition that both the real and imaginary parts of $\Delta_{jk}$ never vanish, which forbids all Fermi arcs, constrains the system to the trivial braid phase.

\begin{figure}
    \centering
    \newcommand{\bx}{\textcolor{lred}{\ensuremath{B_p}}}
    \newcommand{\tbx}{\textcolor{dred}{\ensuremath{\tilde{B}_p}}}
    \newcommand{\bep}{\textcolor{lblu}{\ensuremath{B_\text{EP}}}}
    \newcommand{\tbep}{\textcolor{dblu}{\ensuremath{\tilde{B}_\text{EP}}}}
    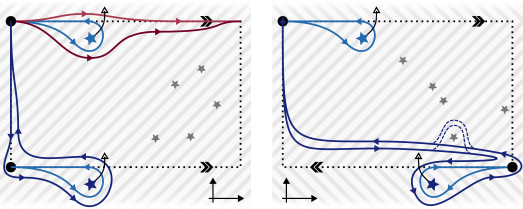
    \caption{
        Rules for braided phase transitions.
        An EP (star marker) with braid invariant \(B_\text{EP}\), measured along the light blue loop, crosses an (a) periodic [(b) anti-periodic] boundary. 
        This leads to a transition from braid \(B_p\) (light red) to \(\tilde B_p = B_\text{EP} B_p\) (dark red).
        The EP's braid invariant changes as well, transitioning according to \cref{eq:phase-transition-rule-ep}. 
        This can be understood by observing that an EP from the [anti-]periodic continuation moves into the fundamental domain as the original EP exits. This entering EP carries the same [inverse] invariant as the exiting EP, measured from the base point in the neighboring domain. 
        The change in invariant to \(\tilde B_\text{EP}\) corresponds to a change of base point associated with the encircling trajectory. 
        If the EP encircles potential other degeneracies (shaded black) before exiting the fundamental domain (dashed blue line in (b)), this rule changes by additional conjugations following the established non-Abelian braiding rules.
        }
    \label{fig:encircling-rules}
\end{figure}

\section{Braid Inversions and Deformations of Exceptional Topology}

Transitions between different gapped phases happen necessarily through gap closing.
Under deformation of $H(p,q)$, pairs of EPs form and continuously move through parameter space. 
If in the process an EP encircles the parameter space, i.e., follows a fundamental loop, this changes some of the fundamental braids, and hence leads to a different gapped phase after all EPs have annihilated pairwise.
This process can also lead to charge inversion of the EP if it goes along the orientation-reversing direction in the parameter space.
If the nonorientable space arises due to a symmetry as in \cref{eq:nonsymmorphic}, this inversion is similar to the inversion of Dirac points in Ref.~\cite{montambauxWindingVectorHow2018}.

\subsection{Phase transition}

We begin by deriving the rule for the phase transition.
As shown in \cref{fig:encircling-rules}, the braids along a $\lambda$-loop in the parameter space before and after the EP crosses it differ precisely by the EP's braid $B_\text{EP}$. 
This result can be seen by noticing the difference between the $\lambda$-loops; a path that passes by an EP on one side differs from the same path passing by on the other side, and that difference amounts to encircling the EP once. 
This implies our first key rule on non-Hermitian phase transitions: an EP crossing a noncontractible $\lambda$-loop changes the corresponding braid as
\begin{equation}
\label{eq:phase-transition-rule-bdy}
    B^{\KB}_{p,q}\to \tilde B^{\KB}_{p,q} = B_\text{EP} B^{\KB}_{p,q},
\end{equation}
for the characterization \(\left(B_p^{\KB}, B_q^{\KB}\right)\) on the Klein bottle. Similar results can be obtained for the real projective plane by replacing \(\left(B_p^{\KB}, B_q^{\KB}\right)\) with $B_{pq}^{\RPtwo}$.
We note again that, as non-Abelian invariants, the explicit braid expression can be different if we choose another base point. 
However, the rule \cref{eq:phase-transition-rule-bdy} is valid for  braid invariants obtained with any common base points in the parameter space, as shown in \cref{fig:encircling-rules}.

\subsection{Braid inversion}

We continue by deriving how the charge of an EP changes \(B_\text{EP}\to \tilde B_\text{EP}\) when it encircles the parameter space along a noncontractible loop.
To make the picture clearer, it is helpful to consider an extended zone scheme, with multiple nonorientable fundamental domains glued together, think of the EP as exiting the fundamental domain into the neighboring unit cell on one side, while a partner EP enters the fundamental domain from the opposite side.
These neighboring unit cells are [mirrored] copies of the fundamental domain if the crossed boundary is [non]orientable (see \cref{fig:encircling-rules}).
Thus, the partner EP that enters the fundamental domain carries the same [inverse] braid charge.
However, this braid is measured from the base point of the neighboring unit cell.
In order to describe the EP properly after it has entered the fundamental domain, we must re-base this invariant, following
\begin{align}
  \label{eq:phase-transition-rule-ep}
  \tilde B_\text{EP} &= B_\text{rebase}\ B_\text{EP}^{\pm1}\ B_\text{rebase}^{-1}, 
\end{align}
where the exponent is \(+1\) if the EP moves in the  orientation-preserving direction and \(-1\) in the orientation-reversing direction.
The conjugation action \(B_\text{rebase}\) is determined by how we measure the EP's charge when it encircles the parameter space (see \cref{fig:encircling-rules}). 
As the EP encircles the parameter space via the directions of the noncontractile loops,
the rebasing action  \(B_\text{rebase}\) is explicitly given by
\begin{equation}
\label{eq:phase-transition-ep-rebase}
    B_\text{rebase} =
    \begin{cases}
        B_{pq}^{\RPtwo} & \text{enc.  \RPtwo},
        \\
        B_p^\KB  & \text{enc. \(\KB\), orientable direction},
        \\
        B_q^\KB B_p^\KB & \text{enc. \(\KB\), nonorientable direction}.
    \end{cases}
\end{equation}
The movement in the orientation-reversing direction inverts and conjugates an EP. We term this encircling operation a non-Abelian charge inversion.

\begin{figure}[htbp]
    \centering
    \newcommand{\lp}{\ensuremath{\lambda}}
    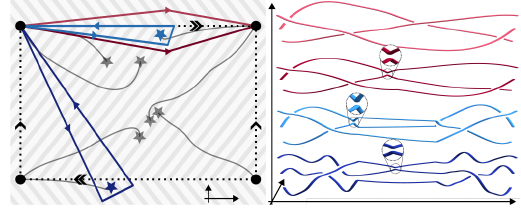
    \caption{
    Phase transition on the Klein bottle. Fermi arcs and eigenvalue braids of the model in \cref{eq:ham-transition}. Increasing parameter \(t\), the gap closes, pairs of EPs form and move through the boundary.
    At critical \(t_0\approx 0.29\), an EP crosses the chosen boundary, changing the fundamental braids from \((B^\KB_p,B^\KB_q) = (\sigma_2\sigma^{-1}_1,\Delta^{-1})\) to \((\sigma^{-1}_1,\Delta^{-1})\).
    (a) The parameter space at \(t=\frac14 < t_0\). 
    Colored lines denote loops that determine the braids relevant to the transition.
    They are piecewise linear instances chosen equivalent to the loops of the same color in \cref{fig:encircling-rules} in terms of encircling. 
    The measured braid is independent of the exact shape or parametrization of the underlying loop.
    Real Fermi arcs (black lines) terminate in EPs (stars). 
    The light blue EP is about to cross the boundary, leaving the fundamental domain while the dark blue EP enters. 
    (b) Spectral braiding along the loops shown in (a) in matching colors, parameterized along arbitrary loop parameter $\lp$. Avoided crossings are highlighted.
    The light blue loop encircles the EP about to cross the boundary, giving rise to its braid charge \(B_\text{EP}=\sigma_1^{-1}\sigma_2^{-1}\sigma_1^{-1}\sigma_2\sigma_1 = \sigma_2^{-1}\).
    The light red path is equivalent to the $p$-boundary before the EP has crossed through, and carries braid
    \(B_p = \sigma_1^{-1}\sigma_2^{-1}\sigma_1\sigma_2 =  \sigma_2\sigma_1^{-1}\).
    The dark red path is equivalent to the $p$-boundary after the EP has crossed through, and carries braid
    \(\tilde B_p = \sigma_1^{-1}\sigma_2^{-1}\sigma_2 = \sigma_1^{-1}\).
    The braid along the boundary changes according to \cref{eq:phase-transition-rule-bdy}.
    The dark blue loop gives the charge of the EP after the boundary crossing,
    \(\tilde B_\text{EP} = \sigma_2 \sigma_1 \sigma_2^{-1}\), in line with \cref{eq:phase-transition-rule-ep,eq:phase-transition-ep-rebase}.
    }
    \label{fig:encircling-nonabelian-numeric}
\end{figure}

\subsection{Deformation: three-band example}
We illustrate such an effect on EPs along a nonorientable direction in a Klein bottle model by performing a linear deformation
\begin{equation}
\label{eq:ham-transition}
H^\KB_\text{transition}(t) = (1-t)H^\KB_\text{gap} + t H^\KB_\text{gapless},
\end{equation}
where 
$H^\KB_\text{gap}$ is defined in \cref{eq:klein-model-gapped} and 
\begin{equation}
\label{eq:klein-gap-closing}
  H^\KB_\text{gapless} = 
  \mqty(
        -a &  1 & 0 \\
        \hphantom{-}1 & b & 1 \\
        \hphantom{-}0 & 1 & a
    ),~\text{where}~
    \begin{cases}
      \,a&\!\!\!\!=i + i e^{2 i q} ,
      \\
      \,b&\!\!\!\!=i (1+\cos{p}) + \sin{p}\cos{q}.
    \end{cases}
\end{equation}
As we show in \cref{fig:encircling-nonabelian-numeric}, the deformation of the exceptional braid topology happens precisely according to the rules in \cref{eq:phase-transition-rule-bdy,eq:phase-transition-rule-ep}.
Tuning away from the gapped model \(H^\KB_\text{gap}\), the gap closes and EPs are pair-created locally. 
One of these carries \(B_\text{EP}=\sigma_2^{-1}\). When this EP crosses the noncontractible loop in the $p$-direction, it changes the eigenvalues braid on the latter from \(B_p=\sigma_2\sigma_1^{-1}\) to \(\tilde B_p = \sigma_1^{-1}\), following \cref{eq:phase-transition-rule-bdy}.
After the EP itself encircles the fundamental domain, its topological charge measured from the nearest loop corner point of the fundamental domain becomes \(\tilde B_\text{EP} = \sigma_2 \sigma_1 \sigma_2^{-1}\).

As the resulting system carries \( \tilde B_q^\KB 
    \tilde B_p^\KB
    \left(\tilde B_q^\KB\right)^{-1} 
   \tilde B_p^\KB = \sigma_2\sigma_1\), the final system is not gapped.
The Abelianization of this expression is $2$, which is consistent with the charge summation rule in Sec.~\ref{sc_not}. 
As gapped phases have zero total Abelianization, in order to transform the system back to a gapped phase again, it is  necessary for the EPs to further encircle the fundamental domain along the $p$-direction.

\section{Gapless Systems and Monopole}\label{sc_epmp}
We now describe gapless systems and what happens when all EPs are brought together.
In situations where summation of the braid charges in a disklike region is nontrivial \( \prod_{\textrm{all EPs}} B_{\textrm{EP}}\ne 1\), the presence of EPs is topologically protected inside the disklike region.
When these EPs are brought to the same position in the fundamental domain, they do not annihilate. Instead, we will have a single multi-level degeneracy carrying the original charge \( \prod_{\textrm{all EPs}} B_{\textrm{EP}}\ne 1\).
The fermion doubling intuition of annihilating them pairwise fails.

The single EPs obtained in this way, as monopoles of eigenvalue braid invariant, are not arbitrary. As we have seen in Sec.~\ref{sc_not}, the disklike region containing all EPs can be chosen to bound the fundamental domain, so the total charge within the disk
must follow \cref{eq_sumchg}. This sum braid is allowed to have nonzero even degree \cref{eq_absumchg}, which goes beyond what is possible in non-Hermitian systems on the torus \cite{yangFermionDoublingTheorems2021,konigBraidprotectedTopologicalBand2023}.

\subsection{Monopole: two-band example}
We illustrate a characteristic monopole of degree two in the two-band gapless \((\sigma_1,1)\)-phase on the Klein bottle with the model
\begin{equation}
\label{eq:model-two-band}
    H_\text{EP}^\KB=
    \mqty(
             -a & 1 \\
             \hphantom{-}1 & a
    ), 
      \,a=-\sin{p}\cos{q} + i (1+\cos p)\left(1+ \ell/2 \cos 2q\right)
\end{equation}
and \(\ell\in[0,1]\) is an adiabatic parameter.
At \(\ell=0\), there are two EPs in this band structure, at \((p,q)=(\pm\pi/2,\pi/2)\).
Both have the same braid invariant \(\sigma_1\), summing to  \(B_\text{tot}^\KB = \sigma_1^2\).
We can fuse them to a single monopole by tuning to \(\ell=1\),
merging the two EPs to a \(\sigma_1^2\)-monopole at \((p,q)=(0,\pi/2)\) (see \cref{fig:branch-cuts}).
This is an unpaired monopole and has total braid degree two, which is forbidden in orientable non-Hermitian systems \cite{konigBraidprotectedTopologicalBand2023}, where the braid degree as an Abelian invariant must vanish.

This single EP also has a characteristic signature in terms of oriented Fermi arcs.
In phases with nonzero total Abelianization, a nonzero number of Fermi arcs must cross a nonorientable boundary.
Since Fermi arcs reverse their direction on such crossing, this adds \(\pm2\) to the total number of outflowing arcs. 
This reflects the evenness requirement for the total Abelianized charge parity in \cref{eq_srab}.

In gapless systems, Fermi arcs can terminate at EPs.
Using the orientation defined in the previous section, open Fermi arcs are oriented away from the most generic twofold EPs with square root dispersion, and towards their inverses, resembling flux lines or Dirac strings.
Whereas a normal EP is the source of one real and one imaginary flux line, the characteristic monopole in the example given in \cref{eq:model-two-band} at \(\ell=1\) is the source of two of each, all oriented away from the EP, corresponding to its Abelianized braid charge \(2\).
Thus, both a nonzero number of total outflowing Fermi arcs, and a nonzero total Abelianization charge necessarily require a nonorientable fundamental domain.

\section{Discussion}
In this manuscript, we have classified exceptional topological phases on two-dimensional nonorientable manifolds.
We have shown that gapped phases on the Klein bottle and real projective plane are determined by torsion and conjugacy problems in the braid group, and that these phases can be distinguished experimentally by their Fermi arcs.

The transitions between distinct gapped phases take place via gapless systems containing EPs, where we have illustrated how telltale Fermi arc signatures arise.
These are closely associated to transitions of the braid when EPs encircle the parameter space in an orientation-preserving or orientation-reversing direction. 
Nonorientability also opens the possibility of gapless phases that violate fermion doubling even in the Abelian limit.
Our work provides insights into how non-Hermiticity and nonorientability interact to produce unique topological phases, expanding the boundaries of non-Hermitian topological band theory, and paving the way to experimental observation of phenomena unique to such combined systems. First implementations exist in photonic, acoustic, and electric systems \cite{xuNonHermitianExceptionalTopology2025,wangAnomalousCollisionExceptional2026,laiObservationExceptionalTopology2026}.

Continuing this line of inquiry, it is known that additional internal symmetries extend and amend the topological classification of non-Hermitian systems \cite{bernardClassificationNonHermitianRandom2002,kawabataSymmetryTopologyNonHermitian2019, yangHomotopySymmetryNonHermitian2024, yoshidaWindingTopologyMultifold2025}, and even allow for the generic occurrence of higher-order EPs \cite{delplaceSymmetryProtectedMultifoldExceptional2021,mandalSymmetryHigherOrderExceptional2021, sayyadRealizingExceptionalPoints2022}. 
An investigation into their interplay with nonorientability would thus be of great future interest.

\begin{acknowledgments} 
JLKK acknowledges insightful discussions with Oscar Arandes, Lukas R\o{}dland, and Tsuneya Yoshida.
AGF, SV, and MS acknowledge support from the U.S. Office of Naval Research (ONR) Multidisciplinary University Research Initiative (MURI) under Grant No. N00014-20-1-2325 on Robust Photonic Materials with Higher-Order Topological Protection.
This material is based upon work also supported in part by the U.S. Army Research Office through the Institute for Soldier Nanotechnologies at MIT, under Collaborative Agreement Number W911NF-23-2-0121, and also in part by the Air Force Office of Scientific Research under the award number FA9550-21-1-0299.
KY is supported by the ANR-DFG project (TWISTGRAPH).
JLKK and EJB were supported by the Swedish Research Council (VR, grant 2018-00313), the Wallenberg Academy Fellows (2018.0460) and Scholars (2023.0256) programs, and the project Dynamic Quantum Matter (2019.0068) of the Knut and Alice Wallenberg Foundation, as well as the Göran Gustafsson Foundation for Research in Natural Sciences and Medicine.
\end{acknowledgments}

%

\clearpage
\section*{Supplemental Material}
\renewcommand{\thefigure}{S\arabic{figure}}
\setcounter{figure}{0}
\subsection{Phase Classification: Generic Surfaces}
We derive the generalized fermion doubling theorem for all possible two-dimensional parameter spaces by an argument analogous to the torus case laid out in Ref.~\cite{wojcikEigenvalueTopologyNonHermitian2022}.

Our starting point is the classification theorem of closed connected two-dimensional manifolds \cite{leeCompactSurfaces2011}.
This classic result in topology states that any closed connected two-dimensional manifold is either the $2$-sphere \(\mathbf{S}^2\), a connected sum of $g$ tori \((\TO)^{\#g}\), or a connected sum of $k$ real projective planes \((\RPtwo)^{\#k}\).
These manifolds all admit a polygonal presentation, which means that they can be obtained from a polygon by identifying sides in a specific pattern.
A standard presentation requires polygons with \(2\), \(4g\), or \(2k\) sides, respectively for the three cases, with edges identified according to the boundary words \(a a^{-1}\) (for \(\mathbf{S}^2\)), \(a_1, b_1, a_1^{-1}, b_1^{-1}, \ldots, a_g, b_g, a_g^{-1}, b_g^{-1}\) (for \((\TO)^{\#g}\)), or \(c_1, c_1, \ldots, c_k, c_k\) (for \((\RPtwo)^{\#k}\)), as we show in \cref{supp:fig:polygonal-presentation}.

\begin{figure}
    \pgfdeclarelayer{background}
    \pgfsetlayers{background,main}
    \pgfkeys{%
      /tikz/on layer/.code={
        \pgfonlayer{#1}\begingroup
        \aftergroup\endpgfonlayer
        \aftergroup\endgroup
      }}
    \centering
    \begin{tikzpicture}
    \tikzset{
        mid arrow/.style={postaction={decorate,decoration={
            markings,
            mark=at position .4 with {\arrow[#1]{stealth}}
          }}},
        dot/.style = {circle, fill=black, minimum size=#1,
              inner sep=0pt, outer sep=0pt},
        dot/.default = 4pt
    }

    \begin{scope}[shift={(-3,0)}, scale=1]
        \node[dot] at (0,-1) (bottom) {}; 
        \coordinate (top) at (0, 1); 
        \draw[thick,mid arrow = black] (bottom) to[bend right=30] (top) ;
        \draw[thick,mid arrow = black] (top) to[bend right=30] (bottom);
        \draw[draw=none,fill=lblu!60!white,on layer=background] (bottom.center) to[bend right=30] (top) to[bend right=30] (bottom.center);
        \node[below=0cm of bottom] {Sphere};

         \node[above right = 0cm and 1mm of bottom] {$a$};
         \node[below left = 0cm and 1mm of top] {$a^{-1}$};
    \end{scope}

    \begin{scope}[shift={(-0.2,0)}, scale=1]
        \foreach \n in {0,...,4}
            {\node[dot,fill=black] (n\n) at (45 * \n - 90:1cm) {};}
        \draw[draw=none,fill=lblu!60!white,on layer=background] (n0.center)
            \foreach \n in {1,...,4} { -- (n\n.center) }
            -- cycle;
        \foreach \n in {1,...,4}
            {\pgfmathtruncatemacro{\m}{\n - 1}
             \draw[thick,mid arrow=black] (n\m) -- (n\n);}
            \path (n0.center) ++(157.5:3mm) node[rotate=-22.5] (d2) {$\cdots$};
            \path (n4.center) ++(202.5:3mm) node[rotate=22.5] (d1) {$\cdots$};
            \path[fill=gray!20!white,on layer=background] (n0.center) -- (n4.center) -- (n4.south west) -- (n0.north west); 
            \shade[left color = white, right color = lblu!60!white,on layer=background] (n0.center) -- (n4.center) -- (d1.center) -- (d2.center) -- cycle;

            \node[above right = 0.5mm and -2mm of n0] {$a_1$};
            \node[above left = 1mm and -2mm of n1] {$b_1$};
            \node[above right = 0cm and -1.5mm of n2] {$a_1^{-1}$};
            \node[above left = 0cm and -3mm of n3] {$b_1^{-1}$};

            \node[below=0cm of n0] {Sum of tori};
        \end{scope}

    \begin{scope}[shift={(3,0)}, scale=1]
         \foreach \n in {0,...,2}
            {\node[dot,fill=black] (n\n) at (90 * \n - 90:1cm) {};}
        \draw[draw=none,fill=lblu!60!white,on layer=background] (n0.center)
            \foreach \n in {1,...,2} { -- (n\n.center) }
            -- cycle;
        \foreach \n in {1,...,2}
            {\pgfmathtruncatemacro{\m}{\n - 1}
             \draw[thick,mid arrow=black] (n\m) -- (n\n);}
            \path (n0.center) ++(135:3mm) node[rotate=-45] (d2) {$\cdots$};
            \path (n2.center) ++(225:3mm) node[rotate=45] (d1) {$\cdots$};
            \path[fill=gray!20!white,on layer=background] (n0.center) -- (n2.center) -- (n2.south west) -- (n0.north west); 
            \shade[left color = white, right color = lblu!60!white,on layer=background] (n0.center) -- (n2.center) -- (d1.center) -- (d2.center) -- cycle;

            \node[above right = 2mm and -2mm of n0] {$c_1$};
            \node[above left = 1mm and -3mm of n1] {$c_1$};
            
            \node[below=0cm of n0] {Sum of projective planes};

    \end{scope}
        \end{tikzpicture}
    \caption{Standard polygonal presentations for closed surfaces. The sides of the polygons are identified following the given boundary word, counterclockwise from a given starting point. 
    The inverse symbol (\(x^{-1}\)) denotes that the edge is attached (to \(x\)) in the direction opposite to the arrow.
    The arrangements of start and endpoints of the edges implies automatically that all corners are identified to the same point (except for the top vertex in the presentation of \(\mathbf{S}^2\).
    The presentations on squares shown in \cref{fig:brillouin-zones} 
    are such polygonal presentations.
    }
    \label{supp:fig:polygonal-presentation}
\end{figure}

Only the third family, the connected sum of projective planes is nonorientable. The cases studied explicitly in the main manuscript correspond to \(k=1\) for \RPtwo{} itself, and \(k=2\) for \KB.
We will in the following consider only the infinite families, as our statements are trivial on the $2$-sphere.

Consider now a Hamiltonian \(H\) defined on such a space \(M\).
EPs are of codimension two \cite{berryPhysicsNonhermitianDegeneracies2004,bergholtzExceptionalTopologyNonHermitian2021}, so in general \(H\) will be degenerate at a set of discrete points \(\Delta \coloneqq \{p_i\in M | i=1,\ldots l\}\) in \(M\).
Restricting the Hamiltonian to avoid these points leads to a well-defined function from the punctured \(M\setminus\Delta\) to the set of \emph{non-degenerate} operators, from which the topological properties arise.

We are therefore trying to topologically classify maps from \(l\)-times punctured \(M\) to the space of non-degenerate operators.
For this argument, we focus on the eigenvalue topology only (which disregards only the usual Chern contribution for gapped phases).
The \(N\) eigenvalues of a non-degenerate Hamiltonian form an unordered tuple of distinct complex numbers, which we regard as a point in the unordered configuration space \(\operatorname{UConf}_N(\C)\).

This \(\operatorname{UConf}_N(\C)\) has the nice topological property of being an Eilenberg--MacLane space of type \(K(\B_N,1)\), meaning it has fundamental group \(\B_n\), and trivial higher homotopy groups.
The braid group on \(N\) strands,  \(\B_N\) arises from the braiding of the \(N\) complex numbers along loops.
Maps into this space are therefore classified precisely by their action on loops, i.e., the space of maps
\begin{equation}
\label{eq:classifying-maps}
    [M\setminus\Delta,\operatorname{UConf}_N(\C)] = \operatorname{Hom}\left(\pi_1(M\setminus\Delta),\B_N\right).    
\end{equation}
is given by the group homomorphisms between the respective fundamental groups.

Our homotopy classification thus concludes by obtaining the remaining fundamental group of these punctured manifolds, for which we turn to the Seifert--van Kampen theorem \cite{leeSeifertVanKampen2011}, which describes how one obtains the fundamental group of a complicated space from a decomposition into simpler parts.
We begin with the punctured manifold in the polygonal representation outlined above, and split it into its interior \(M_i\), and its boundary \(M_b\), chosen to overlap on an annulus \(A = M_i\cap M_b\).

We choose the interior large enough that it contains all of the punctures \(\Delta\). 
It is therefore topologically a \(l\)-times punctured disk, which has as its fundamental group the free group on \(l\) generators (since there is one independent loop around each of the punctures, with no simplifying relations between them), so \(\pi_1(M_i) = \Z^{\star l}\).

We obtain the boundary fundamental group by contracting \(M_b\) to the (one-dimensional) polygon boundary. 
It thus consists of only the non-contractible loops -- for the three classes these are: none, \(a_i\) and \(b_i\), or \(c_i\).
It has as its fundamental group the free group on precisely these \(2g\) or \(k\) generators respectively, since there are again no simplifying relations between the loops;
\begin{equation}
    \pi_1(M\setminus\Delta) = \begin{cases}
        \{1\} & \text{for }M=\mathbf{S}^2,\\
        \Z^{\star 2g} & \text{for }M=(\TO)^{\#g},\text{ or}\\
        \Z^{\star k}  & \text{for }M=(\RPtwo)^{\#k}.
    \end{cases}
\end{equation}

The Seifert-van Kampen theorem states that the fundamental group we are deriving is a so-called amalgamated free product of these two. 
This means it is a free product, with the additional requirement that loops on the intersection annulus \(A\) have to be identified in both groups. 
This annulus can be contracted to a circle, so there is only a single non-trivial loop that we need to consider (the fundamental group is \(\pi_1(A)=\Z\)).

On the interior \(M_i\), this is the loop following the boundary of the disk; we obtain it by encircling every puncture once in the appropriate order.
Similarly on the boundary \(M_B\), encircling the entire polygon once corresponds to this loop along the annulus. 
Identifying the two, we find that the fundamental group \(\pi_1(M\setminus\Delta)\) is
\begin{equation}
    \braket{p_1,a_1, b_1,\ldots p_l,a_g,b_g}{\prod p_i = a_1 b_1 a_1^{-1} b_1^{-1} \cdots a_g b_g a_g^{-1} b_g^{-1}},\text{ or}
\end{equation}
\begin{equation}
    \braket{p_1,c_1,\ldots p_l,c_k}{\prod p_i = c_1 c_1  \cdots c_k c_k}.
\end{equation}
Here we have named the loops around the punctures \(p_i\), and the loops along the sides of the polygon \(a_i, b_i\) or \(c_i\) in a minimal abuse of notation.

This fundamental group is already reminiscent of the sum rule in Eq.~(2) of the main text, and indeed we obtain \cref{eq:classifying-maps} by assigning a braid to each generating loop.
The group relations then give precisely the requirement that the braids around the punctures have to combine to the same total braid as the braids along the boundary.

\emph{Remark}: As noted before, the Klein bottle corresponds to a connected sum of \(k=2\) projective planes. 
The general formalism outlined here leads to a presentation of its fundamental group as \(\braket{a_1 a_2}{a_1^2 a_2^2}\). 
This is a different presentation than \(\braket{a b}{b a b^{-1} a}\) which arises naturally in the setting of our main text.
The two present the same group; their difference amounts to a different choice of fundamental domain as shown, e.g., by the cutting/pasting procedure outlined in Lemma~6.16. of \cite{leeCompactSurfaces2011}.

\begin{table}[htb]
    \centering
    \caption{Braid Sum Rules for Arbitrary Manifolds: We list the independent braids stemming from loops around the manifold, which exist in addition to the braids \(B_{EP_i}\) around each EP. We list the braid sum rule, and its abelianization for each manifold. The group commutator is written as \(\comm{A}{B} = A B A^{-1}B^{-1}\), and \(A_x\) denotes the Abelianization of braid \(B_x\).}
    \begin{tabular}{r|ccc}
        manifold & sphere & orientable \((\TO)^{\#g}\) & non-orientable \((\RPtwo)^{\#k}\)\\
        \hline
        generators &
            ~\(\varnothing\)~ &
            ~\(B_{p_1}, B_{q_1}, \ldots B_{p_g}, B_{q_g}\)~ &
            ~\(B_{p_1}, \ldots,B_{p_k}\)~\\
        \(\prod B_{EP_i} = ~\) &
            1&
            \(\prod_i\comm{B_{p_i}}{B_{q_i}}\)&
            \(\prod_i B_{p_i}^2\)\\
        \(\sum A_{EP_i} = ~\) &
            0&
            0&
            \(2 \sum_i A_{EP_i}\)\\
    \end{tabular}
    \label{tab:my_label}
\end{table}

\subsection{Gapped Phases on \texorpdfstring{\KB}{K2}: Braid Conjugacy}

As shown in the main text, gapped phases on the Klein bottle are indexed by braids that are conjugate to their own inverse (together with a thus conjugating braid) according to Eq.~(5).
The problem reduces to the purely group theoretical search for pairs of braids \((A,B)\) that satisfy
\begin{equation}
    \label{eq:app2:klein-relation}
    A = B A^{-1} B^{-1}.
\end{equation}
We approach this problem by focusing on braid \(A\), for which this equation constitutes a conjugacy requirement.

The structure of this conjugacy problem allows us to make some initial statements.
First, the trivial braid \(A_0=1\) satisfies \cref{eq:app2:klein-relation} for arbitrary conjugating \(B\), leading to the infinite family of gapped phases \((1,B)\cong B_n\).
Since the braid group is torsion-free, the phase \((1,B=1)\) is the only solution with trivial conjugating element \(B=1\).
Second, inversion of \(A\) also inverts its braid degree, while conjugation leaves it invariant, so any solution \(A\) must have zero Abelianization.
As \(\B_2\) (for systems with two bands) is Abelian, \((A_0=1,B \text{ arbitrary})\) constitutes the full set of gapped phases in this case.
Third, for solutions \((A,B)\), arbitrary conjugations \((C A C^{-1}, C B C^{-1})\) by some \(C\in \B_n\), and powers \((A^n,B)\) also solve \cref{eq:app2:klein-relation}.
Fourth, the conjugation \(B\) can be multiplied by elements in \(\{A\}\)'s centralizer. 
In particular, \(B\) can always be multiplied by arbitrary powers of \(\Delta^2\) (which generates the braid group's center). 
Taken together, these four statements already allow us to construct various infinite families of gapped phases. 

Listing all gapped phases remains non-trivial, since \cref{eq:app2:klein-relation} is not well-behaved under group multiplication: products of solutions \(A_1*A_2\) are not necessarily solutions. 
In particular, the first solution found in the main text, \(A_1 = \sigma_2\sigma_1^{-1}\), multiplied by another solution in its same conjugacy class gives the first counterexample \(\tilde A_1 = A_1 * (\sigma_2\sigma_1^2) A_1 (\sigma_2\sigma_1^2)^{-1}\), which does not satisfy \cref{eq:app2:klein-relation} as we show at the end of this appendix.
The general question of conjugacy in braid groups is understood but in general quite involved.

In the following we provide an explicit solution to the conjugacy problem in \(\B_3\) and use it to construct all solutions to \cref{eq:app2:klein-relation}.
This illustrates the richness of gapped phases on \KB{} already in systems with three bands.
We show in particular that not all braids with zero Abelianization are conjugate to their inverse.
We use known conjugacy results inherited from \emph{Garside} groups, a class of groups containing the braid groups \cite{gonzalez-menesesBasicResultsBraid2011,garsideBraidGroupOther1969}.
Our method generalizes to arbitrary \(\B_n\), where it is however much more involved owing to the greater complexity of braids with more strands.

Our argument proceeds as follows: We describe the solution of the word problem, the so-called \emph{left normal form} in which any braid can be expressed uniquely \cite{elrifaiAlgorithmsPositiveBraids1994}.
We then solve the conjugacy problem in a formulation that is more appropriate to the task at hand than previous solutions \cite{truolUpsilonInvariant12023, vilchezPositive3braidsKhovanov2025}.
We provide a simple description of conjugacy classes in terms of binary sequences, a unique representative for every class, and a prescription for obtaining conjugating braids to other representatives.
We continue by showing how conjugacy classes transform under inversion, which leads us to a complete classification of solutions to \cref{eq:app2:klein-relation}.
We conclude with examples of solutions, as well as interesting non-solutions that showcase the complexity of the solution set.

Any braid may be written uniquely as an arbitrary power of the fundamental braid \(\Delta = \sigma_1\sigma_2\sigma_1\), times a positive braid (containing no inverses \(\sigma_i^{-1}\)).
This works since an inverse generator \(\sigma_{1,2}^{-1}\) can be written as \(\Delta^{-1}*\sigma_{1,2}\sigma_{2,1}\), and occurrences of the fundamental braid can be moved to the left via \(\sigma_{1,2} \Delta = \Delta \sigma_{2,1}\)
This positive braid is made unique by absorbing all further powers of \(\Delta\) into the prefactor in this way.
To identify the remaining positive braids uniquely (avoiding ambiguities like  \(\sigma_1\sigma_3 = \sigma_3\sigma_1\), which is not an issue on \(\B_3\)), their constituents are grouped into so-called \emph{simple elements}.
These are all short positive braids that can be completed to form \(\Delta\) by right-multiplication. 
The unique \emph{left normal} form is then
\begin{equation}
\label{eq:app2:left-normal-form}
    \Delta^l a_1 \ldots a_r
\end{equation}
where \(a_i\) are simple, and arranged such that \(a_ia_{i+1}\) is not simple.

In \(\B_3\), the simple elements are the atoms \(\sigma_{1,2}\) (which we will call \emph{short}), together with \(\sigma_1\sigma_2\) and \(\sigma_2\sigma_1\) (which we will call \emph{long}).
Allowed sequences of simple elements in our case require simple elements that end in \(\sigma_{i}\) to be succeeded by elements that begin in the same \(\sigma_{i}\). 
The long elements therefore constitute the only allowed switch between sequences of \(\sigma_1\) and sequences of \(\sigma_2\).

For any element in the sequence \(\{b_i\}\), there is one unique short, and one unique long element that may succeed it. 
This simplifies our search for conjugacy classes satisfying \cref{eq:app2:klein-relation}: Conjugation of a braid by \(\Delta\) exchanges \(\sigma_1 \rightleftharpoons \sigma_2\), but conserves the short/long distinction. 
Instead of sequences of simple elements, we can therefore specify conjugacy classes as
\begin{equation}
\label{eq:app2:left-normal-form-conjugate}
     \Delta^l b_1 \ldots b_r
\end{equation}
with \(\{b_i\}\) an arbitrary binary sequence of \(\{\text{short},\text{long}\}\).
A standard representative of a given class is obtained by starting the left normal form in \cref{eq:app2:left-normal-form} with \(\sigma_1\) or \(\sigma_1\sigma_2\), depending on \(b_1\). 

We now use the conjugation search algorithm for Garside groups \cite{birmanConjugacyGarsideGroups2007,gebhardtSolvingConjugacyProblem2010} to find and distinguish all conjugacy classes on \(\B_3\).
For any braid, there is a conjugation operation called \emph{cyclic sliding}, which does not increase the number \(r\) of simple elements in its reduced positive braid \cite{gebhardtCyclicSlidingOperation2010}. 
Thus, repeated cyclic sliding must lead to an attractor cycle of braids of a fixed length, and the conjugacy problem can be investigated on such cycles only.
For braids on three strands, it turns out that all these cycles are of length one (the braids are then called \emph{rigid}) -- with the exception of three special classes we discuss at the end of this paragraph.
By direct computation, we find that cyclic sliding shortens all braids whose exponent \(l\) of \(\Delta\) and number of occurrences of \(\text{long}\) elements have different parity.
For the inclined reader, we provide the following simplified recipe which works equivalently on \(\B_3\): 
Conjugate braids that end in \(\sigma_i\) by the other generator (\(\Delta^{-1} \sigma_i \Delta\)), bring to left normal form, and repeat.
The aforementioned exceptions consist of the classes that cannot be shortened: 
braids \(\Delta^n\) with no positive braid, and those with only one simple element.
The former are rigid and behave like the families we have already found.
Those braids with a single simple element are not rigid, but form \(2\)-cycles
\(\Delta^{\text{even}}\sigma_1\sigma_2 \rightleftharpoons \Delta^{\text{even}}\sigma_2\sigma_1\), and \(\Delta^{\text{odd}}\sigma_1 \rightleftharpoons \Delta^{\text{odd}}\sigma_2\) 
with the respective other simple element under cyclic sliding or conjugation by \(\Delta\).
These exceptional families of conjugacy classes are of no interest to the problem at hand, as \(\Delta^k\) inverts to \(\Delta^{-k}\) and only the trivial class is a solution, while the other two families exchange place under inversion and thus contain no solutions.

In order to construct all classes, we note that arbitrary positive braids can be made rigid if prefixed by appropriate powers of \(\Delta\), namely even powers if they start and end on the same atom, and odd powers otherwise. This exhausts the regular conjugacy classes.

To find all conjugacy classes, we must next identify which classes in this exhaustive list are equivalent, for which we now investigate the action of conjugation on rigid braids.
As mentioned earlier, conjugation by \(\Delta\) swaps the group's two generators. 
By explicit calculation, we find that conjugation by the group's generators \(\sigma_i\) precisely accesses the cyclic permutations of a positive braid's sequence \(\{b_i\}\) in the presentation of \cref{eq:app2:left-normal-form-conjugate}, after another reduction through cyclic sliding.
The conjugacy classes can thus be conveniently indexed by binary sequences in \(\{\text{short},\text{long}\}\) modulo cyclic permutations, which exhaustively and uniquely determines all conjugacy classes (save for the aforementioned exceptions).

Finding the conjugating element between two braids in the same class -- the conjugacy search problem -- is solved by reducing the respective braids to their rigid representatives via cyclic sliding, and then implementing cyclic permutations via an appropriate conjugation.

We are now ready to tackle inversion. 
In a given positive braid, inversion exchanges short for long elements like \(\sigma_1 \overset{-1}{\rightleftharpoons} \left(\Delta^{-1}\right) \sigma_1\sigma_2\), and vice versa, as well as inverting the ordering of elements. 
For rigid braids, the inverse does not simplify further, beyond having to commute all occurrences of \(\Delta\) to the left, hence it is also rigid.
The conjugacy class identified by a given \(\Delta^p\), and a sequence in \(\{\text{short},\text{long}\}\) of length \(l\), inverts to the conjugacy class \(\Delta^{-p-l}\) and the reversed, and \(\text{short} \rightleftharpoons\text{long}\) flipped sequence.

The conjugacy classes that solve \cref{eq:app2:klein-relation} are precisely the fixed points under this operation.
They are given by those binary sequences \(\{b_i\}\) that under binary negation and sequence inversion are mapped to a cyclic permutation, prefixed by the appropriate power of \(\Delta\).
Constructing all such classes is straightforward.
We note that they must contain an equal number \(m\) of short and long elements, and that the exponent of the fundamental braid is also fixed to \(\Delta^{-m}\) by the zero braid degree requirement.

A simpler way to represent these classes is by commuting the instances of \(\Delta\) back into the braid, and reverting the original replacement of inverse generators for all long elements. In this way, we find that all long elements turn into the same inverse generator, so the solutions can be represented by words in \(\{\sigma_1,\sigma_2^{-1}\}\). They solve \cref{eq:app2:klein-relation} precisely if this sequence, under reverting the order and replacing \(\sigma_1 \leftrightharpoons \sigma_2^{-1}\), turns into a cyclic permutation of its original form.

We conclude by providing some examples of increasing complexity. 
Sorting by length \(2 m\) of the sequence \(\{b_i\}\), the first conjugacy class of braids satisfying \cref{eq:app2:klein-relation} is given by 
\begin{equation}
    \Delta^{-1} \text{short} * \text{long} \ni A_1 =\Delta^{-1} \sigma_1 (\sigma_1\sigma_2) = \sigma_2\sigma_1^{-1},
\end{equation}
containing precisely the braid used in Figure~2~(a),
which we know to be conjugate to its inverse by \(B_1 = \Delta^\text{odd}\).
The next shortest example is at \(m=2\),
\begin{equation}
     \Delta^{-2} \text{short}^2 * \text{long}^2 \ni A_2 
     =\Delta^{-2} \sigma_1^2 (\sigma_1\sigma_2) (\sigma_2\sigma_1) = \sigma_1^{2}\sigma_2^{-2},
\end{equation}
which is again conjugate to its inverse by \(\Delta\).
Using the formalism outlined here, it is not hard to show that every conjugacy class satisfying \cref{eq:app2:klein-relation} must contain an element that is conjugate to its inverse by \(\Delta\).
The other element at length \(m=2\) is also a solution, it corresponds to \(A_1^2\).

Finally, we provide some counterexamples; braids that are not conjugate to their inverse. 
First, of course, are all braids with nonzero degree.
Second, among the braids with zero degree is
\begin{equation}
    \Delta^{-2}\text{short}^6 \overset{-1}{\rightleftharpoons} \Delta^{-4} \text{long}^6, 
\end{equation}
a counterexample (which we learned from  J. Gonz\'ales-Meneses) as it contains an unequal number of short and long elements.
As noted earlier, \(A_1 * c_{\sigma_2\sigma_1^2}(A_1)\) lies in this class, showing explicitly that products of solutions need not be solutions.
Third and finally, at \(m=4\), we find
\begin{equation}
    \Delta^{-4} \text{short}^3 \text{long}^2 \text{short}^1\text{long}^2 \ni \sigma_1^3\sigma_2^{-2}\sigma_1\sigma_2^{-2},
\end{equation}
the simplest conjugacy class that does not solve \cref{eq:app2:klein-relation} (this is clear by inspection of the sequence) even though it has zero Abelianization and equally many occurrences of short and long elements, i.e., can be represented as a word in \(\{\sigma_1,\sigma_2^{-1}\}\).

\end{document}